\documentstyle[prd,aps]{revtex}
\begin{document}
\draft
\twocolumn[\hsize\textwidth\columnwidth\hsize\csname
@twocolumnfalse\endcsname
\preprint{SUSSEX-AST 96/6-1, astro-ph/9606047}
\title{The spectrum of curvature perturbations from hybrid inflation}
\author{Juan Garc\'\i a-Bellido$^\ast$ and
David Wands$^{\dagger\ast}$}
\address{$^\ast$Astronomy Centre, University of Sussex, Falmer,
Brighton BN1 9QH,\ U.K.\\
$^\dagger$School of Mathematical Studies, University of Portsmouth,
Portsmouth PO1 2EG,\ U.K.}
\date{\today}
\maketitle
\begin{abstract}
We study the amplitude and spectral tilt of density perturbations in
the simplest hybrid inflation models. We give an exact expression for
the amplitude of quantum fluctuations on all scales in the limit where
we can neglect the backreaction on the metric. This is a very good
approximation for values of the inflaton field well below the Planck
scale and our results remain valid far from the usual massless limit.
We confirm that the primordial density spectrum in this model has a
constant spectral index $n>1$ over all observable scales. For the
small values of the tilt ($n<1.4$) required by observations, the
results remain close to those obtained using the quasi-massless
approximation.
\end{abstract}
\pacs{PACS numbers: 98.80.Cq \hspace*{1mm} Preprint
SUSSEX-AST 96/6-1,  RCG 96/08,  astro-ph/9606047}

 \vskip2pc]

Inflation provides a compelling explanation for many aspects of the
observed universe, but it is the spectrum of primordial density
perturbations that provides the best hope of distinguishing between
different possible models of inflation.  Vacuum fluctuations of the
inflaton field during inflation can be swept up to astrophysical
scales by the rapid expansion. The spectrum of perturbations on
super-horizon scales are usually computed using the Bunch-Davies
vacuum for a massless field~\cite{BD} for scales within the horizon
and matching to an approximate homogeneous solution outside that
horizon. Exact solutions to the equations of motion for linear
perturbations about the homogeneous field are only known in a handful
of special cases~\cite{LS,SL,Easther}.

One particular model of inflation that has received special
attention recently is hybrid inflation~\cite{hybrid,LL93,CLLSW}.  It
is possible to find models of this kind in
supersymmetric~\cite{CLLSW,Lazarides,SO(10)} and some
supergravity~\cite{CLLSW,Stewart} particle physics models. The
possibility that particle physics may give a workable model of
inflation is very attractive. In particular, the mass scales present
in the model appear naturally in hidden-sector supersymmetry
breaking~\cite{Guth}. These models have a very rich low energy
phenomenology, as well as important cosmological and astrophysical
implications, some of which were explored in Ref.~\cite{GBLW}. For
certain values of the parameters, the model may have a second stage of
inflation, which may lead to the production of cosmologically
interesting black holes and topological defects. As part of our
calculation of the density perturbations present in these two-stage
models of hybrid inflation, we found solutions to the equations of
motion for linear perturbations of both scalar fields when their
backreaction on the metric can be neglected~\cite{GBLW}.

In this paper we show that our result for a single field evolving
during hybrid inflation also applies when there is only one stage of
inflation, and is an (almost) exact solution for a wide range of
parameters in the hybrid inflation scenario, well beyond the usual
slow-roll approximation. In fact, corrections to the amplitude and
tilt of curvature perturbations produced during inflation can become
large away from the quasi-massless limit. In some models of hybrid
inflation it is possible to have a significant positive tilt; however,
the maximum value of the tilt allowed by observations is
$n\lesssim1.4$, where deviations from the quasi-massless results
remain small.

The simplest realization of chaotic hybrid inflation is provided by
the potential~\cite{hybrid}
\begin{equation}\label{hybrid}
V(\phi,\psi) = \left(M^2-{\sqrt{\lambda}\over2}\psi^2 \right)^2
 + {1\over2}m^2\phi^2 + {1\over2}\gamma\,\phi^2\psi^2 \, .
\end{equation}
The equations of motion for the homogeneous fields are then
\begin{eqnarray}\label{EQM}
\ddot\phi + 3H\dot\phi
 &=& - (m^2 + \gamma\psi^2) \phi\, , \\
\ddot\psi + 3H\dot\psi
 &=& (2\sqrt\lambda M^2 - \gamma\phi^2 - \lambda\psi^2) \psi \, ,
\end{eqnarray}
subject to the Friedmann constraint
\begin{equation}
H^2 = {8\pi\over3M^2_{\rm P}} \left[ V(\phi,\psi)
 + {1\over2}\dot\phi^2 + {1\over2}\dot\psi^2 \right] \, .
\end{equation}

The potential has a local minimum with respect to the field $\psi$ at
$\psi=0$ for $\phi^2>\phi_c^2$, where
\begin{equation}
\phi_c^2 \equiv {2\sqrt{\lambda} \over \gamma} \, M^2 \, .
\end{equation}
Thus inflation can occur while $\psi=0$ and the potential
simply reduces to
\begin{equation}\label{potential}
V(\phi) = M^4 + {1\over2}m^2\phi^2 \,.
\end{equation}
while $\phi$ rolls down the potential towards $\phi_c$.
For a wide range of
parameters the constant term always dominates on scales of interest
and the Hubble expansion can be taken to be de Sitter expansion with
$H=H_0\equiv\sqrt{8\pi/3}\,M^2/M_{\rm P}$.
It is then useful to write the bare masses of the two fields $\phi$ and
$\psi$ relative to the Hubble scale as
\begin{equation}\label{AB}
\alpha \equiv {m^2\over H_0^2} = {3\over8\pi}\,
{m^2 M_{\rm P}^2 \over M^4}\, ,
\end{equation}
and
\begin{equation}
\beta \equiv 2\sqrt\lambda \, {M^2\over H_0^2} =
{3\over4\pi}\, {\sqrt{\lambda}M_{\rm P}^2\over M^2}\, ,
\end{equation}
respectively.

The detailed form of the potential involving $\psi$ is unimportant,
beyond ensuring that $\psi=0$ is a stable minimum for
$\phi^2>\phi_c^2$, but becomes unstable for $\phi^2<\phi_c^2$. We will
restrict our analysis here to the case where inflation effectively
ends once $\phi$ reaches $\phi_c$. For couplings $\lambda$ and
$\gamma$ not very much less than unity in Eq.~(\ref{hybrid}) the phase
transition from the false vacuum ($\psi=0$) to the true vacuum
($\psi^2\simeq 2M^2/\sqrt{\lambda}$) always occurs rapidly, bringing
inflation to an end in less than a Hubble time. The case of small
couplings has recently been considered in Ref.~\cite{Guth,GBLW} and
shown to require $\beta\gg1$, to ensure that the duration of inflation
after $\phi=\phi_c$ is short enough to prevent too many large black
holes to be formed.

In the case of a single scalar field $\phi$ evolving during inflation,
one usually resorts to the slow-roll, $\dot\phi^2\ll V(\phi)$, and
quasi-massless, $V''\ll H^2$, approximations to make analytic
progress. This allows one to reduce the equations of motion for the
scalar field to a first-order equation. However in our case we wish to
consider values of $\alpha$ not much below unity so the quasi-massless
approximation may not be very good. Instead we will assume that the
potential energy, and hence the Hubble rate, remain constant which
allows us to integrate the full second-order equation of motion for
$\phi$. This amounts to neglecting the backreaction of the $\phi$
field upon the background spacetime. Thus we require
\begin{equation}\label{backreact}
{1\over2} m^2 \phi^2 + {1\over2} \dot\phi^2 \ll M^4 \, .
\end{equation}
We will see that this condition is easily satisfied for a wide
range of parameters.

For $\phi>\phi_c$, the $\psi$ field remains trapped in the stable
minimum at $\psi=0$ and we have effectively single-field inflation. The
$\phi$ field's mass is constant and the equation of motion becomes
\begin{equation}
\phi'' - 3\phi' + \alpha \phi = 0 \, ,
\end{equation}
where a prime denotes a derivative with respect to $N=H_0 (t_e-t)$,
the number of $e$-folds to the end of inflation.\footnote{Note
that $N$ is defined here as a positive quantity which decreases to
zero at the end of inflation.} This can be readily integrated to give
\begin{eqnarray}\label{r}
\phi(N) & = & \phi_+ \exp(r_+N) + \phi_- \exp(r_-N) \,,\nonumber\\
r_\pm & = & {3\over2} \mp \sqrt{{9\over4}-\alpha}\,.
\end{eqnarray}
For $\alpha>9/4$ the solution describes damped oscillations about
$\phi=0$, while for $\alpha<9/4$ the asymptotic solution is $\phi =
\phi_c \exp (r N)$ where $r\equiv r_+>0$, which approaches the slow-roll
solution $\phi = \phi_c \exp (\alpha N/3)$ for $\alpha\ll1$.

Note that the condition for neglecting the field's backreaction,
Eq.~(\ref{backreact}), then becomes
\begin{equation}
\label{back2}
4\pi\, r\, \phi^2 \ll M_{\rm P}^2 \, .
\end{equation}
Thus our analysis remains valid even for $\alpha>1$ as long as we
consider values of the field $\phi$ much less than the Planck scale.

In the quasi-massless approximation, the amplitude of quantum
fluctuations of the field at horizon crossing ($k=aH$) is taken to
be $H/2\pi$. However, if the mass of the $\phi$ field is not necessarily
much smaller than the Hubble scale, corrections to this massless field
result could be large.

On the other hand, for small $\phi\ll M_{\rm P}$ we can neglect the
gravitational backreaction of the field, see Eq.~(\ref{back2}). The
equation of motion for linear perturbations in $\phi$ can then be written
as
\begin{equation}
\ddot{\delta\phi} + 3H \dot{\delta\phi} + \left({k^2\over a^2} +
\alpha H^2\right)\delta\phi =0 \, .
\end{equation}
Note that when $\psi=0$, the evolution of $\delta\psi$ and $\delta\phi$
decouple. We can write this equation, in terms of the canonically
quantized field~\cite{Mukhanov} $u\equiv a\delta\phi$, as
\begin{equation}\label{uv}
u_k'' + \left(k^2 - {2 - \alpha\over\eta^2}\right) u_k = 0\,,
\end{equation}
where primes denote derivatives with respect to conformal time,
$\eta=-1/aH$, and we have chosen $\eta=-1$ when $\phi=\phi_c$.

Since the mass of the $\phi$ field is constant, we can write an exact
expression for the quantum fluctuations,~\cite{BD,SL}
\begin{equation}
u_k(\eta) = {\sqrt\pi\over2\sqrt k}\,e^{i(1-r)\pi/2}\,
(-k\eta)^{1/2}\,H_{3/2-r}^{(1)}(-k\eta)\,,
\end{equation}
where $r=r_+$ is defined in Eq.~(\ref{r}). This has been normalized to
have the correct flat-space behavior
\begin{equation}
u_k \to {e^{ik\eta} \over \sqrt{2k}} \, ,
\end{equation}
as $-k\eta\to\infty$, while as $\phi\to0$, and $-k\eta\to0$, we find
\begin{equation}
\label{asymv}
u_k(\eta) = {C(r)\over\sqrt{2k}}\,e^{i(1-r)\pi/2}\,
(-k\eta)^{r-1}\,,
\end{equation}
where
\begin{equation}\label{Cr}
C(r) = 2^{-r}\,{\Gamma(3/2-r)\over\Gamma(3/2)}\,,
\end{equation}
as shown in Fig.~1. 

We will write the power spectrum of any quantity $A$ as
$\,{\cal P}_A \equiv (k^3/2\pi^2) \langle |A|^2 \rangle$.
Equation~(\ref{asymv}) thus gives a scale-invariant spectrum of the
growing-mode perturbations at horizon crossing, $k\eta_*=~-1$,
\begin{equation}\label{deltaphi}
{\cal P}^{1/2}_{\delta\phi_*} = C(r)\,{H\over2\pi}\,.
\end{equation}
Note that the coefficient $C(r)$ gives a constant correction
(independent of scale) to the usual amplitude of perturbations
(obtained in the slow-roll limit where $C(0)=1$). Near $\alpha=9/4$,
or $r=3/2$, the coefficient $C(r)$ becomes large, see Fig.~1, giving a
significant amplification of the perturbations compared to the usual
slow-roll approximation.

A similar expression was obtained in Ref.~\cite{SL} for natural
inflation in the small angle approximation, where the potential is
approximately given by Eq.~(\ref{potential}) with $m^2\to-m^2$ and
thus $\alpha\to-\alpha$ and $r\to-3/2+\sqrt{9/4+|\alpha|}>0$. In
natural inflation the validity of both the small angle approximation
and neglecting the backreaction, Eq.~(\ref{backreact}), become worse
and must eventually fail as the field approaches the end of inflation,
although this will only affect small scales. In our case the
validity of the approximation improves as the field evolves towards
the end of inflation.

Quantum fluctuations of the scalar field are responsible for curvature
perturbations on comoving hypersurfaces, which can be evaluated as the
change in the time (or number of $e$-folds) it takes to end inflation.
In the case of adiabatic perturbations, e.g. in single-field inflation
driven by the field $\phi$, the amplitude of the curvature perturbation
on comoving hypersurfaces remains fixed on super-horizon
scales, so it can be calculated as $\delta N =
[H\delta\phi/\dot\phi]_*$ at horizon crossing, where $\delta\phi_*$ is
given by Eq.~(\ref{deltaphi}).

At large values of $\phi\gg\phi_c$ the $\psi$ field has a large positive
mass-squared and remains fixed at $\psi=0$. The amplitude of $\psi$
fluctuations crossing outside the horizon are negligible. Thus we need
only consider adiabatic fluctuations $\delta\phi_*$ along the
trajectory, giving an amplitude of perturbations
\begin{equation}\label{dNphi}
{\cal P}^{1/2}_{\delta N}  = {C(r) \, H\over2\pi \,r\,\phi_*}\, .
\end{equation}
The power spectrum of curvature perturbations on comoving
hypersurfaces, ${\cal R} = \delta N$, is then given by
\begin{equation}\label{PR}
{\cal P}_{\cal R}(N) = {C(r)^2\over4\pi^2r^2}\,
{\gamma\over\beta}\,e^{-2rN}\,,
\end{equation}
where $N$ is the number of $e$-folds to the end of inflation, which
occurs at $\phi=\phi_c$. If these curvature perturbations are
responsible for the observed temperature anisotropies in the microwave
background, Eq.~(\ref{PR}) gives a constraint on the parameters of the
model.  The low multipoles of the angular power spectrum measured by
COBE~\cite{COBE} gives a value ${\cal P}_{\cal R}(N_{\rm CMB}) \simeq
3\times 10^{-9}$, on the scale of our current horizon.

Because the comoving scale at horizon crossing is just proportional to
the scale factor, $k \propto e^{-N}$, the scale dependence of the
power spectrum in Eq.~(\ref{PR}) readily gives the tilt of the
spectrum as
\begin{equation}
n-1 \equiv {d\ln{\cal P}_{\cal R}\over d\ln k} = 2r \, ,
\end{equation}
which becomes $2\alpha/3$ in the slow-roll limit. Note that the tilt
is always positive, giving a ``blue'' perturbation
spectrum~\cite{hybrid,LL93,CLLSW,Jim,Mollerach}.

This is one of the few cases in inflationary cosmology where we have
an (almost) exact expression for the amplitude of curvature
perturbations.  The only approximation we have made is to assume that
the energy density remains constant, see Eq.~(\ref{backreact}), so
that we can neglect the backreaction on the metric.  Using
Eqs.~(\ref{back2}) and~(\ref{dNphi}), we see that this will be true as
long as
\begin{equation}\label{back3}
{8C(r)^2\over3r}\,{M^4\over M_{\rm P}^4} \ll {\cal P}_{\cal R}\, .
\end{equation}
This constraint is shown in Fig.~2. Expanding 
$C(r)$ given in Eq.~(\ref{Cr}), it follows that the allowed range of
$r$, for $M^4/M_{\rm P}^4 \ll {\cal P}_{\cal R}$, is
\begin{eqnarray}
r &\gg& \, {8\over3{\cal P}_{\cal R}}\,{M^4\over M_{\rm P}^4}\,,\nonumber\\
\left({3\over2} - r\right)^2 &\gg& 
 \, {8\over9\pi{\cal P}_{\cal R}}\,{M^4\over M_{\rm P}^4}
\,.
\end{eqnarray}
For ${\cal P}_{\cal R}\simeq3\times 10^{-9}$ and $M\ll
10^{16}$~GeV the right-hand-sides of these constraints are so small
that $r$ is effectively left as a free parameter between zero and
$3/2$, corresponding to $\alpha$ in the range $0<\alpha<9/4$. Thus in
principle for this model one could have any value of the tilt in the
range $1<n<4$.

Observations of the microwave background impose a direct constraint 
on the spectral index of the curvature perturbations.  Present limits
from the lowest multipoles of the microwave sky give
$n=1.2\pm0.3$~\cite{COBE} at the $1\sigma$ level, corresponding to
$\alpha\leq0.75$.  This implies that the size of the correction
coefficient in Eq.~(\ref{PR}) lies in the range $0.86\leq C\leq1$. A
precise measurement of $n$, which would be possible with the next
generation of satellite experiments~\cite{PSI}, would give a much
tighter constraint on $\alpha$.

There is another important constraint on the allowed values of the
tilt of the spectrum from small scales. The microwave background
observations constrain the amplitude of perturbations to be small on
large scales, but a positive tilt leads to an increasing amplitude of
curvature perturbations on smaller scales. The maximum amplitude
depends on both the tilt and the number of $e$-folds to the end of
inflation. Large amplitude perturbations on small scales may lead to
the production of primordial black holes~\cite{Jim}.  The most
conservative constraint is to require ${\cal P}_{\cal R}(N)<1$ on all
scales. Combined with the observed value of ${\cal P}_{\cal R}$ on our
present horizon scale, this implies
\begin{equation}\label{PBH}
2r\, (N_{\rm CMB}-N) < - \ln{\cal P}_{\cal R}(N_{\rm CMB}) \, .
\end{equation}
The precise value of $N_{\rm CMB}$ and thus the maximum tilt is
dependent upon factors such as the reheat temperature $T_{\rm rh}$,
\begin{equation}\label{NCMB}
N_{\rm CMB} \simeq 46 +{2\over3}\,\ln\left( {M\over10^{11}{\rm GeV}}
\right) + {1\over3}\,\ln\left({T_{\rm rh}\over10^7{\rm GeV}}\right)\,.
\end{equation}
Taking $N_{\rm CMB}=46$ and $N=0$ in Eq.~(\ref{PBH}) requires $r<0.2$
and $\,n<1.4$.  However in some models of hybrid inflation it is
possible to have a second stage of inflation~\cite{Guth,GBLW} after
the phase transition at $\phi=\phi_c$ and thus the minimum value of
$N$ is greater than zero.  This would give a weaker constraint on $r$.

Possibly the most robust present constraint comes from combining the
anisotropies on large scales with the absence of spectral distortions
in the microwave background's black-body spectrum which would be
produced by the damping of large perturbations on intermediate scales
during the radiation dominated era~\cite{Daly}. This implies that we
should require $n<1.56$ at a 95\% confidence level~\cite{HSS}.
Combining the COBE normalization with observations of large-scale
structure may provide an even stronger constraint of
$n<1.25$~\cite{Pedro}, but this is much more model dependent since the
growth of large-scale structure is sensitive to various other
cosmological parameters~\cite{Lucchin,Pedro}.

Possible values of the parameters, assuming $N_{\rm CMB}<46$, are, for
example
\begin{equation}
\alpha = 0.6 \,, \hspace{1cm} \beta =100\,, \hspace{1cm} \gamma =60\,.
\end{equation}
Note that the energy scale during inflation, $M$, can take any value
below $10^{16}$~GeV.  For $M = 10^{-8}\,M_{\rm P} \simeq 10^{11}~{\rm
GeV}$, we have $m\simeq2$~TeV and
$m_\psi\equiv\sqrt{2}\,\lambda^{1/4}M\simeq30$~TeV in
Eq.~(\ref{hybrid}). These are very natural parameters from the point
of view of supersymmetry. The scalar fields could correspond to flat
directions, e.g. moduli fields, which get mass corrections of order
the gravitino mass $m_{3/2}\sim1$~TeV when supersymmetry is broken in
the observable sector, see Ref.~\cite{Guth,GBLW}. For these
parameters, the tilt of the primordial spectrum of density
perturbations becomes $n=1.4$. Note also that the phase transition
that triggers the end of inflation occurs rapidly for these
parameters, and we are able to avoid a second stage of inflation and
the danger of producing too many large black holes, as discussed in
Ref.~\cite{GBLW}.

Note that in Ref.~\cite{CLLSW} an upper limit on $n<1.14$ was obtained
for values of the dimensionless couplings $\gamma$ and $\lambda$ not
far from unity. On the other hand, we are able to obtain a larger tilt
by considering arbitrary values of the couplings. For instance the
value $\beta=100$ given above corresponds to a very small coupling,
$\lambda\simeq10^{-27}$. However such flat potentials are quite
natural in supersymmetric models, see Ref.~\cite{Guth,GBLW}. The
absolute upper limit $n<1.3$ given in~\cite{CLLSW} comes from taking
$N_{\rm CMB}=60$ in Eq.~(\ref{PBH}), which corresponds to 
$M\sim10^{16}$~GeV and $T_{\rm rh}\sim10^{13}$~GeV in Eq.~(\ref{NCMB}),
much greater than the values considered above.

For completeness we note that the amplitude of the gravitational wave
spectrum can also be computed,
\begin{equation}
{\cal P}_{\rm g} = {16\over\pi}\,{H_0^2\over M_{\rm P}^2}
 = {128\over3} \, {M^4 \over M_{\rm P}^4} \, .
\end{equation}
A small Hubble constant during inflation implies
that the contribution of gravitational waves to the microwave
background anisotropies will be small~\cite{GW,LL93,CLLSW}.
Indeed the validity of our curvature perturbation calculation, see
Eq.~(\ref{back3}), requires their contribution to be negligible.

In summary, we have shown that hybrid inflation may be added to a
select group of inflationary models~\cite{LS,SL,Easther} for which the
curvature perturbation spectrum may be calculated (almost)
exactly. The spectral index of the power spectrum is $n=1+2r$ where
$r\equiv3/2-\sqrt{9/4-m^2/H^2}$, which approaches the quasi-massless
approximation $n_{\rm sr}\simeq1+2m^2/3H^2$ for $m\ll H$, see Fig.~3.
The correction coefficient for the amplitude of perturbations for a
given value of the field $\phi$ at horizon crossing, relative to the
quasi-massless result, is given by $C(r)$ in Eq.~(\ref{Cr}), which is
never far from unity for $m<H$, see Fig.~1. Nonetheless, our
calculation remains valid for a wide range of parameters, not
necessarily close to the usual quasi-massless limit. In principle one
can obtain large values for the tilt of the spectrum, but in practice
the maximum value is constrained by observations. 

It is worth emphasizing that these hybrid inflation models contain
mass scales that are consistent with natural values present in
particle physics models~\cite{hybrid,CLLSW,Guth,GBLW}. Perhaps it will
be possible in the not-so-far future to test this models both by
cosmological observations of the microwave background and in high
energy particle physics experiments.

\subsection*{Acknowledgements}
The authors thank Andrei Linde, Andrew Liddle and David Lyth for their
helpful comments. They are also grateful to the Physics Department at
Stanford University, where this work was done, for their
hospitality. The authors acknowledge support from PPARC (U.K.) and
from a NATO Collaborative Research Grant, Ref.~CRG.950760.

\

\begin{center}
{\bf Figure captions}
\end{center}

\noindent
{\bf Figure 1:} The coefficient $C(r)$ in Eq.~(\ref{Cr}), giving the
amplitude of perturbations relative to the massless field limit as a
function of the parameter $r$.

\vspace{5mm}
\noindent
{\bf Figure 2:} The maximum value of $\log_{10}(M/M_{\rm P})$ given in
Eq.~(\ref{back3}) for ${\cal P}_{\cal R}=3\times10^{-9}$ as a function
of the parameter $r$. The allowed region is below the curve, which
effectively means that, for $M\ll 10^{-3} M_{\rm P}$, any value of $r$
in the range $(0,3/2)$ is in principle possible.

\vspace{5mm}
\noindent
{\bf Figure 3:} The spectral index, $n$, of the curvature
perturbations as a function of $m/H$. The dashed line shows the tilt
obtained using the quasi-massless approximation, $n\simeq 1+2m^2/3H^2$.

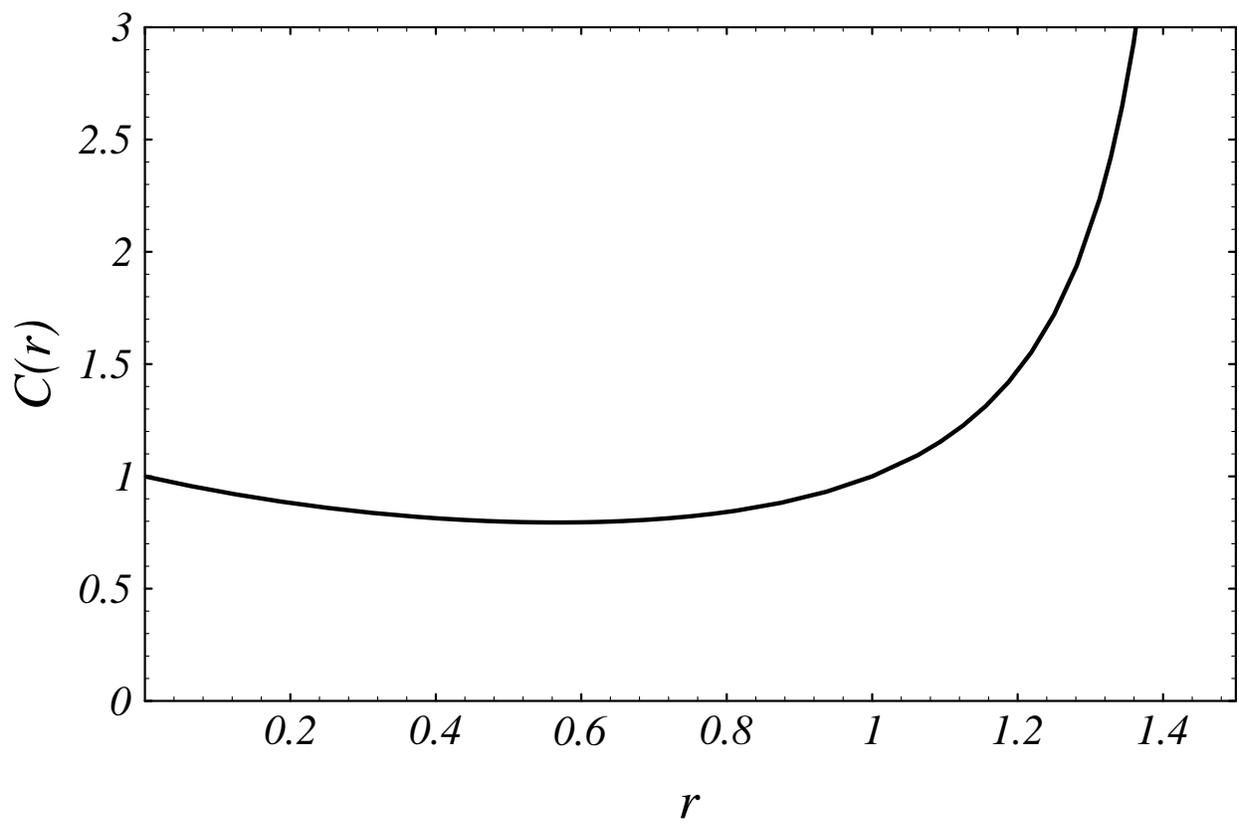
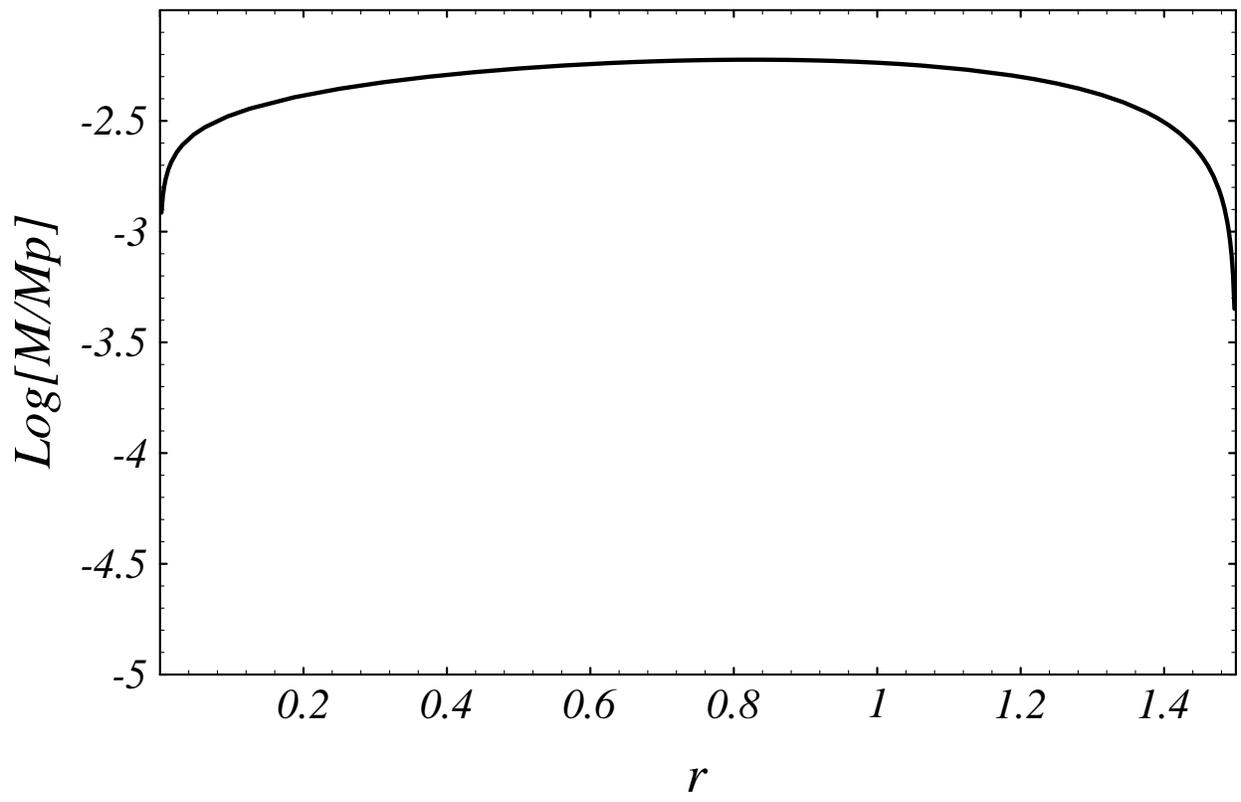
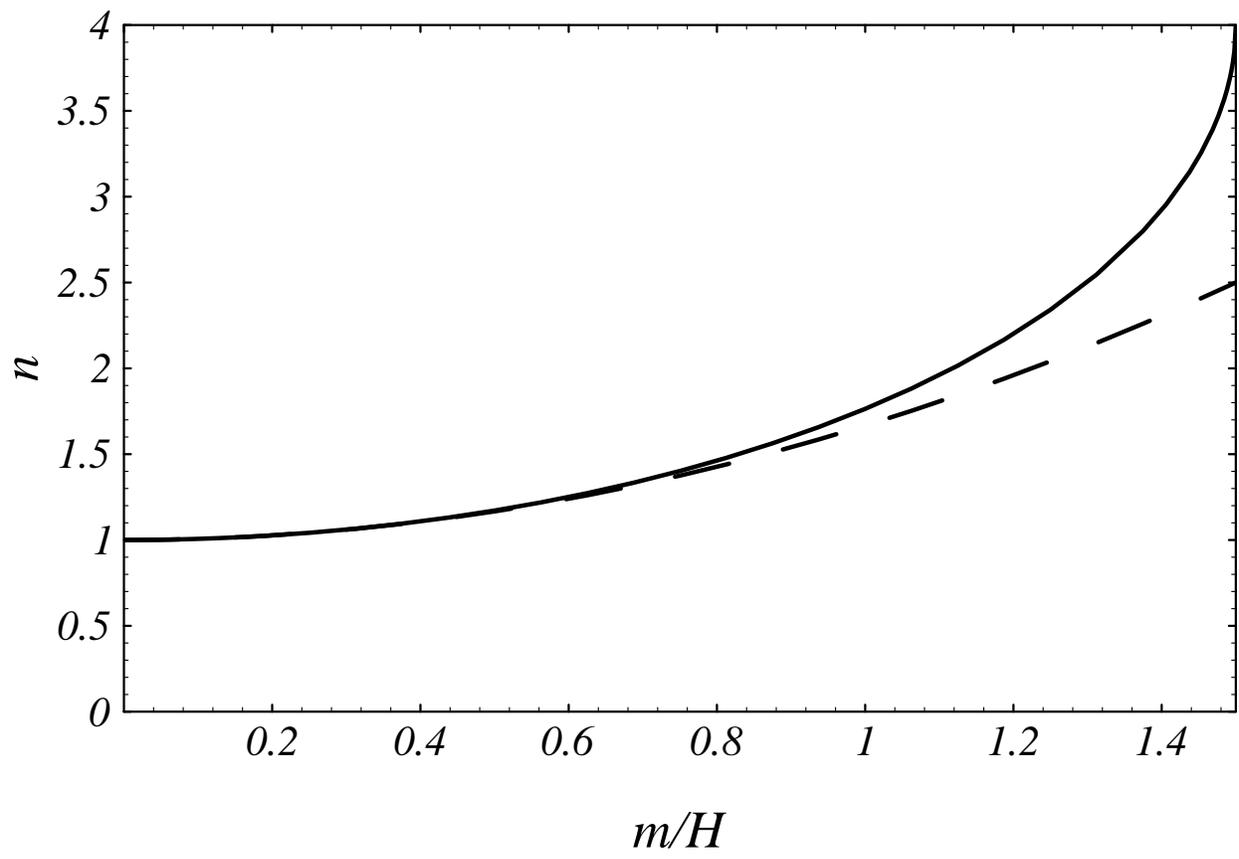
\end{document}